%% file: main.tex
\documentclass[aps,prl,twocolumn,amsmath,amssymb,floatfix,longbibliography,superscriptaddress]{revtex4-2}

\usepackage{color}
\usepackage{mathrsfs}
\usepackage{graphicx}
\usepackage{dcolumn}
\usepackage{bm}
\usepackage{empheq}
\usepackage[hidelinks]{hyperref}
\usepackage{xcolor}
\hypersetup{
    colorlinks,
    linkcolor={blue},
    citecolor={blue!80!black},
    urlcolor={blue!80!black}
}
\usepackage[normalem]{ulem}
\usepackage{amsmath,amsfonts,amssymb,ulem}
\usepackage{epstopdf}
\usepackage{xcolor}
\usepackage[separate-uncertainty]{siunitx}
\usepackage{upgreek}
\newcommand{\re}[1]{{\color{black}#1}}

\usepackage{graphicx}

\begin{document}

\title{Laser wakefield acceleration of ions with a transverse flying focus}

\author{Zheng Gong}
\email[]{zgong92@itp.ac.cn}
\affiliation{Department of Mechanical Engineering, Stanford University, Stanford, California 94305, USA}
\affiliation{Institute of Theoretical Physics, Chinese Academy of Sciences, Beijing 100190, China}
\author{Sida Cao}
\affiliation{Department of Mechanical Engineering, Stanford University, Stanford, California 94305, USA}
\author{John P. Palastro}
\affiliation{Laboratory for Laser Energetics, University of Rochester, Rochester, New York 14623, USA}
\author{Matthew R. Edwards}
\email[]{mredwards@stanford.edu}
\affiliation{Department of Mechanical Engineering, Stanford University, Stanford, California 94305, USA}

\date{\today}
\begin{abstract}

The extreme electric fields created in high-intensity laser-plasma interactions could generate energetic ions far more compactly than traditional accelerators. 
Despite this promise, laser-plasma accelerator experiments have been limited to maximum ion energies of $\sim$100\,MeV/nucleon.
The central challenge is the low charge-to-mass ratio of ions, which has precluded one of the most successful approaches used for electrons: laser wakefield acceleration.
Here we show that a laser pulse with a focal spot that moves transverse to the laser propagation direction enables wakefield acceleration of ions to GeV energies in underdense plasma.
Three-dimensional particle-in-cell simulations demonstrate that this relativistic-intensity ``transverse flying focus" can trap ions in a comoving electrostatic pocket, producing a monoenergetic collimated ion beam.
With a peak intensity of $10^{20}$\,W/cm$^2$ and an acceleration distance of 0.44 cm, we observe a proton beam with 23.1 pC charge, 1.6 GeV peak energy, and 3.7\% relative energy spread. 
This approach allows for compact high-repetition-rate production of high-energy ions, highlighting the capability of more generalized spatio-temporal pulse shaping to address open problems in plasma physics.

\end{abstract}
\maketitle
Beams of energetic ions have applications ranging from particle \cite{PhysRevLett.131.101801} and nuclear physics \cite{kondo2023first} to laboratory astrophysics \cite{marcowith2016microphysics}, high-energy-density science \cite{schaeffer2022proton}, and medicine \cite{schardt2010heavy_tumor}. 
The electric fields of standard ion accelerators are limited to 100 MeV/m by radio-frequency breakdown \cite{wiedemann2015particle}, so relativistic ion energies require hundreds of meters of acceleration distance.
In contrast, plasmas can support acceleration gradients three orders-of-magnitude higher ($>$100 GeV/m) \cite{Daido_review_ion,macchi_2013_RMP}, motivating efforts to build advanced compact ion accelerators based on laser-driven plasmas. The standard approach to laser ion acceleration is target normal sheath acceleration (TNSA) \cite{mora2003plasma_TNSA}, where a high-intensity laser pulse striking the front surface of a thin solid-density target drives a stream of electrons through the target, setting up an electrostatic ion-accelerating sheath on the rear surface.
At higher laser intensities, 
laser hole boring \cite{pukhov1997laser_HB,macchi2005laser,robinson2009relativistically,naumova2009hole}, 
collisionless shock acceleration \cite{silva2004proton_shock,ji2008generating,fiuza2012laser_shock}, 
light sail acceleration \cite{esirkepov2004highly,klimo2008monoenergetic,yan2008generating,qiao2009stable,chen2009enhanced_PRL}, 
breakout afterburner acceleration \cite{yin2011three}, magnetic vortex acceleration \cite{nakamura2010high_MVA,bulanov2010generation_MVA}, and controlling optical tweezer acceleration via the beat pattern of two colliding lasers~\cite{wan2020ion}, can also accelerate ions from solids. 
Although separating electrons and ions in a solid density plasmas produces large fields, the spatial extent  is small, and once an ion overtakes the localized potential, its energy accumulation terminates.
Additionally, solid targets impose practical issues, including low repetition rates and sensitivity to laser temporal contrast \cite{kaluza2004influence}. 
As a result, maximum ion energies from plasma-based accelerators are limited at $\sim$100 MeV/nucleon  \cite{clark2000measurements_18MeV,snavely2000intense_58MeV,maksimchuk2000forward_2MeV,hegelich2006laser,toncian2006ultrafast,willingale2006collimated,henig2009radiation_CP,haberberger2012collisionless,kar2012ion,bin2015ion_CP_LP,palaniyappan2015efficient,wagner2016maximum_85MeV,zhang2017collisionless,scullion2017polarization_25MeV,higginson2018near_94MeV,ma2019laser,mcilvenny2021selective,wang2021super,rehwald2023ultra,dover2023enhanced,martin2024narrow,ziegler2024acceleration},
and laser-ion accelerators are not yet suitable for applications like radiotherapy, which require $>$200 MeV/nucleon \cite{schardt2010heavy_tumor}.

In this Letter, we propose an alternative method for ion acceleration enabled by a novel form of spatiotemporal control.
This new type of laser-ion accelerator combines the advantages of laser wakefield acceleration (LWFA) with a transversely moving flying focus (TFF). \re{As shown by the detailed comparison between LWFA-TFF and existing laser-plasma ion-acceleration approaches in the Supplemental Material (SM)~\cite{Supplemental_Material}}, the proposed mechanism uses moderate laser intensities ($10^{20}$ W/cm$^2$) and extended acceleration distances (mm-cm) to accelerate ions to up to GeV energies, far surpassing what other approaches can achieve with similar laser intensity.
LWFA \cite{tajima1979laser,mangles2004monoenergetic,geddes2004high,faure2004laser,RMP_LWFA_esarey_2009}, where an electrostatic field forms behind a laser pulse travelling through underdense plasma, is an attractive particle acceleration mechanism: electron LWFA has been demonstrated with high repetition rates, $>$8 GeV electron energies \cite{gonsalves2019petawatt,aniculaesei2024acceleration}, and monoergetic spectra~\cite{wang2021free}.
Applying the success of LWFA to ions, however, is challenging, because ions are far too heavy to gain relativistic energy from the oscillating laser field \cite{salamin2008direct} or the electrostatic field~\cite{liu2022accelerating} within the duration of a laser pulse. 
Ion LWFA requires bridging relativistic laser propagation and sub-relativistic initial ion speeds, via, for example, near-critical plasma densities to reduce the laser group velocity \cite{brantov2016synchronized,liu2020front}---which requires experimentally challenging plasma profiles---or laser intensities sufficiently high that ions reach relativistic velocities within the pulse duration \cite{shen2007bubble}. This second approach needs laser peak powers above $\sim$300 PW, far beyond current capability. 

Instead, here we match the ion and laser pulse velocities using recently-developed optical techniques for spatiotemporal control, e.g.,\ the flying focus  \cite{sainte2017controlling,froula2018spatiotemporal}. A flying focus utilizes chromatic focusing of chirped pulses \cite{sainte2017controlling,froula2018spatiotemporal}, spatially dependent pulse delays \cite{liberman2024use,pigeon2024ultrabroadband}, or fast adjustment of an optic's refractive index \cite{simpson2022spatiotemporal}, to control the trajectory of a focal point over distances far larger than a Rayleigh range. This control offers enormous advantages for plasma Raman amplification \cite{turnbull2018raman}, photon acceleration \cite{howard2019photon}, dephasingless electron acceleration \cite{palastro2020dephasingless,caizergues2020phase}, table-top x-ray lasers \cite{kabacinski2023spatio}, and strong field quantum electrodynamics \cite{formanek2024signatures}.
A key benefit of flying focus approaches in plasma physics is that the most complicated components are the optics, allowing for simpler plasma mechanisms.
Unfortunately, for subrelativistic flying-focus velocities, the finite Rayleigh length makes light field gradients in the longitudinal direction small~\cite{ramsey2023exact}, preventing the formation of the strong, slow-moving electrostatic fields required for ion acceleration, as shown in the SM~\cite{Supplemental_Material}.
To avoid this, we use transverse motion of the laser focal spot to combine a sharp intensity gradient with tunable sub-relativistic focal spot velocity, allowing the efficient acceleration of ions from rest in underdense plasma.


\begin{figure}
\includegraphics[width=0.48\textwidth]{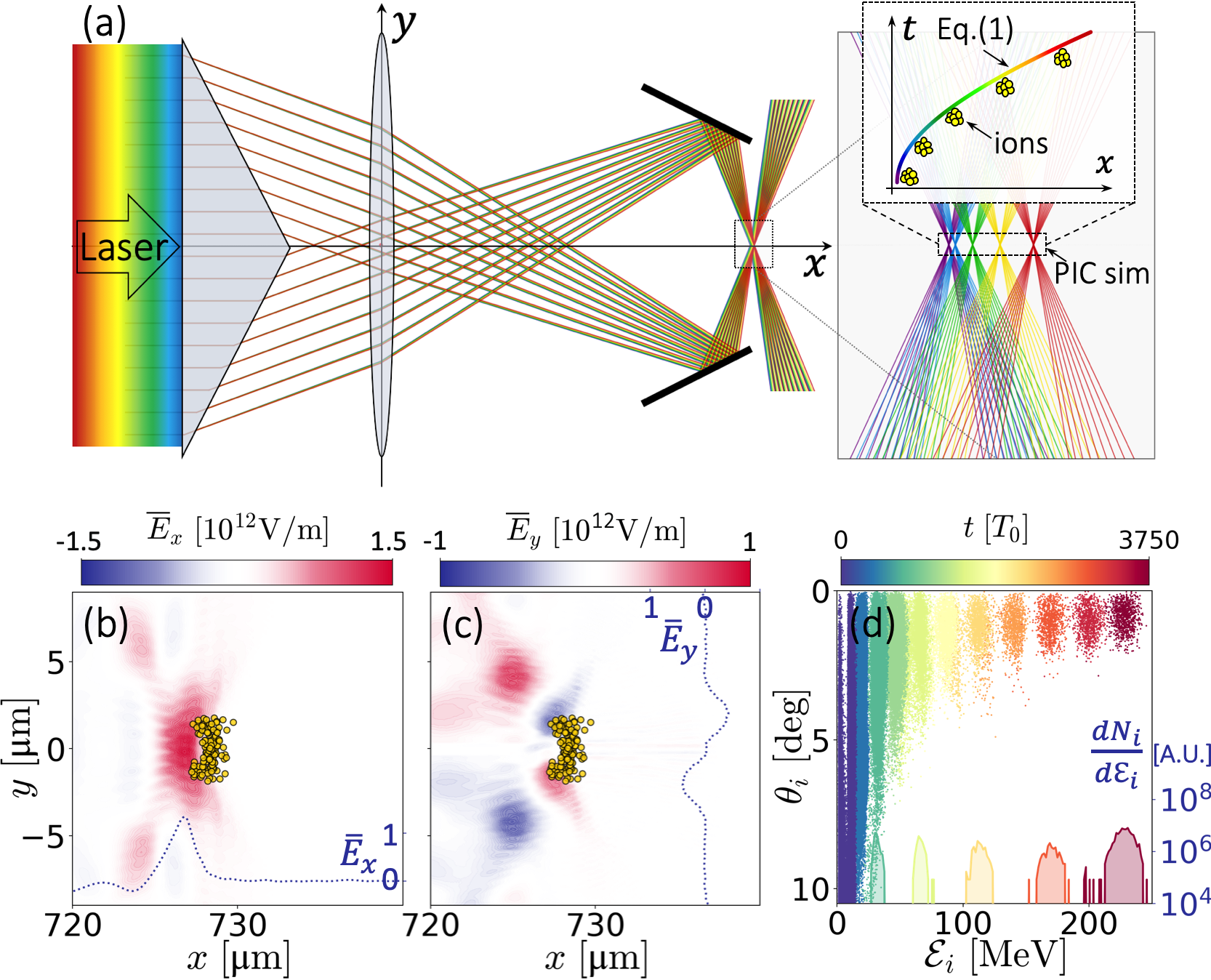}
\caption{(a) Schematic of an axisymmetric transverse flying focus, where solid lines represent light ray propagation and colors denote frequency components. The panel at right shows details near focus. PIC simulations are conducted within the dashed black box, where the inset shows flying focus position as given by Eq.~\eqref{eq:xf_synchronized}.
(b)-(d) 3D PIC simulations. 
Distribution of (b) $\overline{E}_x$ and (c) $\overline{E}_y$ at $t=3150T_0$. In (b)(c), yellow dots show trapped protons with electric fields shown by blue dotted lines. (d) The proton distribution in energy-angle $(\mathcal{E}_i, \theta_i)$ space for different times.}
\label{fig:PIC_2d_two_pulse}
\end{figure}

Consider the focal spot of a laser pulse whose envelope has a velocity perpendicular to the laser propagation direction, as might be created, for example, by a chirped pulse passing though a grating followed by a lens. 
Unlike a stationary focal spot in an underdense plasma, where the strong transverse electrostatic field created by the ponderomotive expulsion of electrons accelerates ions to approximately 10 MeV over microns \cite{macchi2009electric}, a moving focal spot can keep ions within the accelerating field over an extended distance.
An axisymmetric system with transverse focal spot motion can be constructed from a combination of axicon and dispersive optics: e.g., an axicon lens, spherical lens, and axicon mirror as drawn in Fig.~\ref{fig:PIC_2d_two_pulse}a.
The axicon and spherical lens alone would produce a ring focus, with the chromatic dispersion of these two optics creating rings of slightly different radii for each color. When combined with an axicon mirror, the ring foci can be made to collapse onto the $x$ axis. 
Then, if the lens and mirror parameters are chosen appropriately, each color will focus at a desired location along the $x$-axis~\cite{animation_fig2a_schematic,Supplemental_Material}.
Apart from this chromatic approach, echelons or optics with time-dependent refractive indices are alternatives to realize the transverse flying focus in experiments.
This focusing geometry produces TeV/m accelerating fields in the $x$ direction and ion-focusing fields in the $y$ and $z$ directions, as shown by three-dimensional (3D) particle-in-cell (PIC) simulations in Fig.~\ref{fig:PIC_2d_two_pulse}bc.
If the focus and resulting field move along an appropriate trajectory, ions can be accelerated indefinitely in the TeV/m field, allowing for energy accumulation limited only by the energy of the laser pulse.

\begin{figure}
\centering
\includegraphics[width=0.46\textwidth]{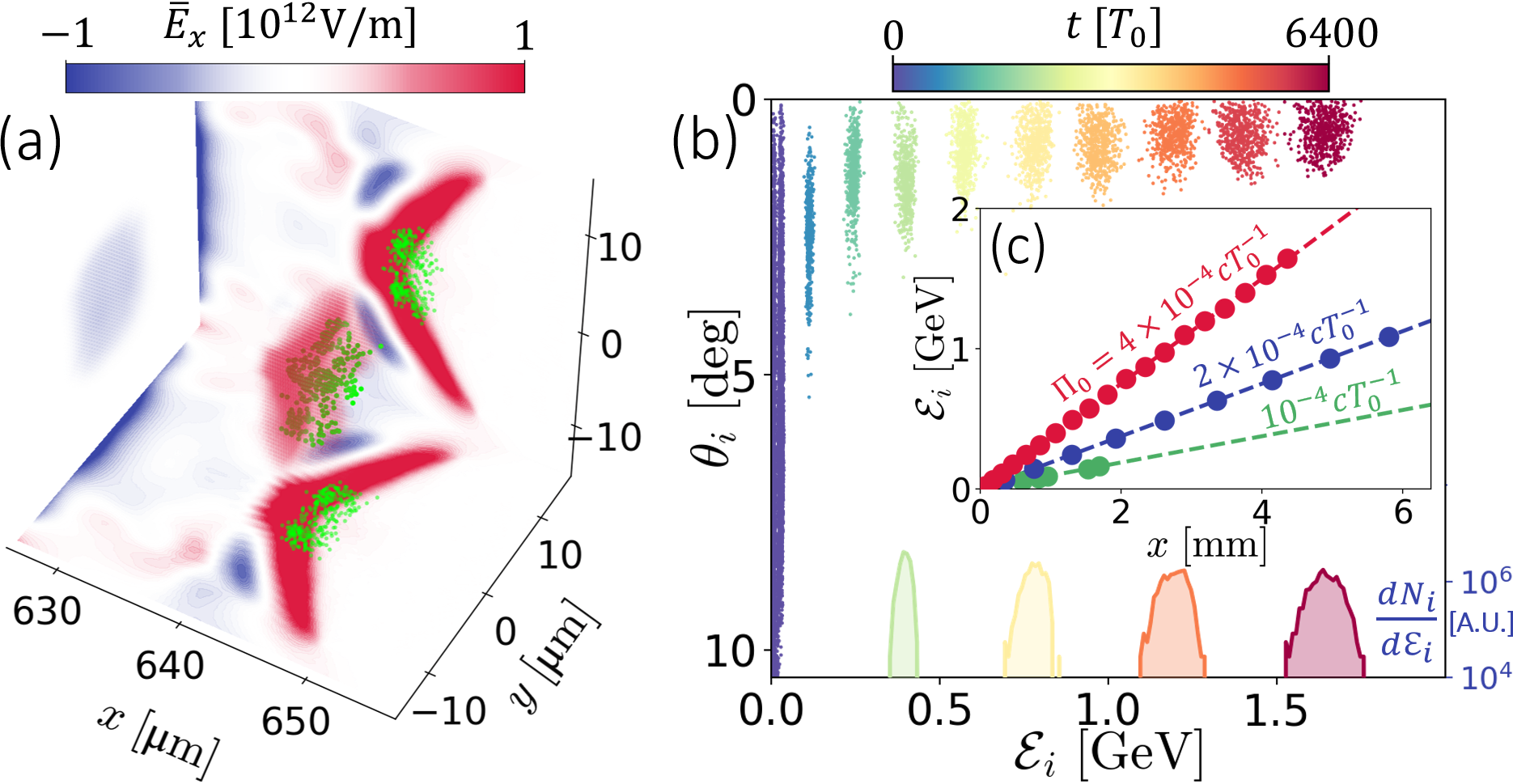}
\caption{ (a) Spatial distribution of the accelerating field $\overline{E}_x$ at $t=6.4\,$ps, where dots represent accelerated protons.
(b) Proton divergence distribution and energy spectra $dN_i/d\mathcal{E}_i$. 
(c) Dependence of proton energy $\mathcal{E}_i$ on the acceleration length $x$ for $\Pi_0=4\times10^{-4}$, $2\times10^{-4}$, and $10^{-4}cT_0^{-1}$, where circles show simulation results and the dashed lines indicate analytic predictions.}
\label{fig:PIC_3d_en_theta}
\end{figure}

\begin{figure*}
\centering
\includegraphics[width=0.96\textwidth]{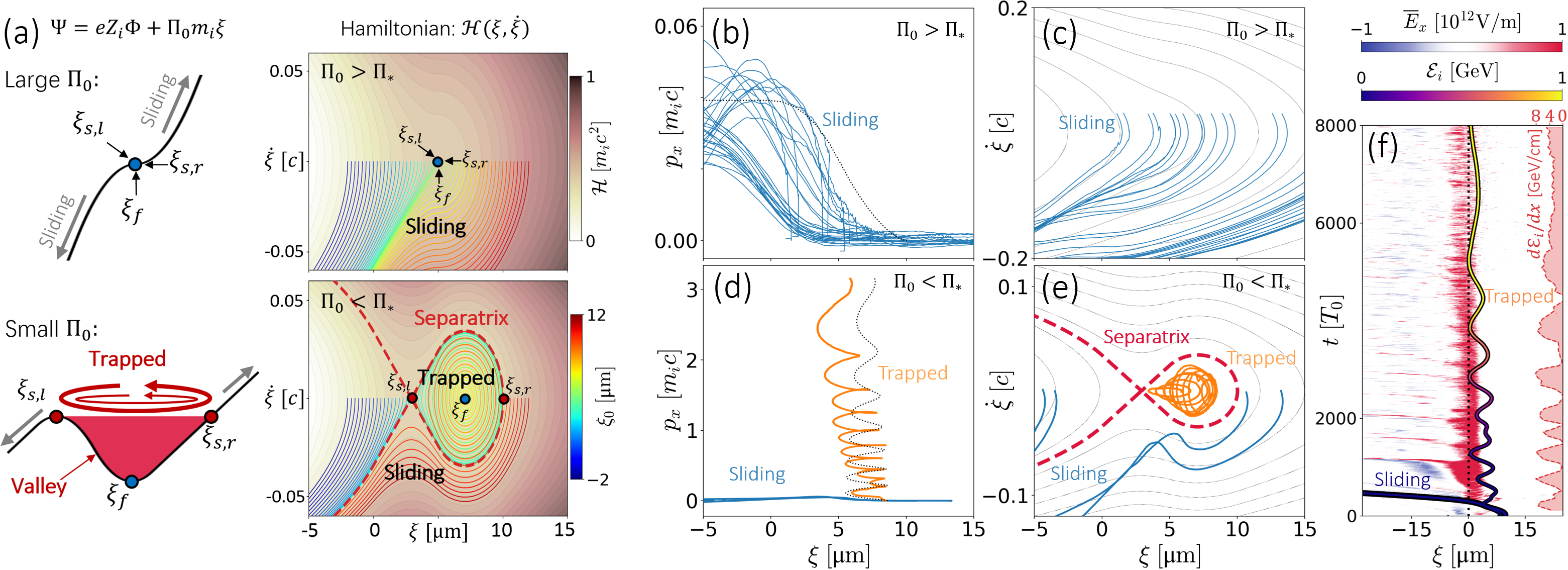}
\caption{(a) Electric potential $eZ_i\Phi$ combined with either a large or small inertial potential $\Pi_0m_i\xi$ produces total potentials $\Psi=eZ_i\Phi+\Pi_0m_i\xi$ and Hamiltonians $\mathcal{H}(\xi,\dot{\xi})$ associated with sliding (upper panels, larger acceleration $\Pi_0$) or trapping (lower panels, smaller acceleration $\Pi_0$). 
Dashed red lines mark the separatrix that divides the trapping and sliding regimes. Solid lines show ion evolution in $(\xi,\dot{\xi})$ space with different initial position $\xi_0$.
(b)-(e) Particle tracks from 2D PIC simulations, where orange and blue lines represent typical trapped and sliding protons, respectively. Black dotted lines in (b)(d) show analytically predicted trajectories and in (c)(e) render the contours of $\mathcal{H}$. (b)(c) and (d)(e) correspond to $\Pi_0=1.2\times10^{-3}cT_0^{-1}$ and $\Pi_0=4\times10^{-4}cT_0^{-1}$, respectively.
(f) Trajectory of the trapped proton in (e) with respect to the accelerating field $\overline{E}_x$ (at the slice $y=0$) in the co-moving frame, where the red shadow gives the instantaneous acceleration gradient.}
\label{fig:Hami_inertial}
\end{figure*}

For efficient ion energy gain, the location of the accelerating field maximum in time ($x_f(t)$)  should match the trajectory of an ion accelerated by a constant force to relativistic velocity: 
\begin{equation}\label{eq:xf_synchronized}
x_f=\frac{c^2}{\Pi_0}\left[\sqrt{\left(\frac{\Pi_0 t}{c}\right)^2+1}-1\right]
\end{equation}
where $\Pi_0$ is the apparent acceleration of the field maximum position and $c$ is the speed of light.
The ions will accelerate based on their mass ($m_i$), charge ($e Z_i$), and the field strength ($E_x$), so this matching requires $\Pi_0 = e Z_i E_x/m_i$.
PIC results shown in Fig.~\ref{fig:PIC_2d_two_pulse}d demonstrate that an ion bunch can be accelerated to more than 200 MeV for a focal spot at position $x_f$ given by Eq.~\ref{eq:xf_synchronized} (see Fig.~\ref{fig:PIC_2d_two_pulse}a).
In these simulations, the laser pulse was focused to a peak intensity of $1.1\times10^{20}$ W/cm$^2$ in a hydrogen plasma with $n_e = 0.05n_c$, where \re{$n_c\equiv\omega_0^2m_e/(4\pi e^2)$}, $m_e$ is the electron mass, $e$ is the elementary charge, and $\omega_0$ is the laser frequency. The acceleration of the focus was $\Pi_0=2\times10^{-4}cT_0^{-1}$, where $T_0\approx3.3$ fs is the laser period [see SM for detailed parameters \cite{Supplemental_Material}]. 
In both 2D and 3D simulations, this produced a stable moving electric field $\overline{E}_x$ (Fig.~\ref{fig:PIC_2d_two_pulse}b) and corresponding transverse field $\overline{E}_y$ that suppressed the protons' transverse dispersion (Fig.~\ref{fig:PIC_2d_two_pulse}c).
The proton distributions shown in Fig.~\ref{fig:PIC_2d_two_pulse}d demonstrate that the acceleration process yielded a monoenergetic proton beam with peak energy $\mathcal{E}_{peak}\approx 230\,$MeV, where $\mathcal{E}_i\equiv (\gamma_i-1)m_ic^2$ is the ion kinetic energy, $\gamma_i$ is the relativistic Lorentz factor, and $\theta_i$ is the ion divergence half-angle.
The ions were accelerated over 1$\,$mm, resulting in energy gains that are almost two orders of magnitude larger than proton energies achievable in a static transverse field. 
The proton beam was collimated with an energy spread $\delta \mathcal{E}_i/\mathcal{E}_{peak}\approx 3.9\%$ and an angular spread $\delta \theta_i \approx 0.89^\circ$ after an acceleration distance of $\delta x\approx0.1\,$cm (Fig.~\ref{fig:PIC_2d_two_pulse}d).

Over longer acceleration distances, this mechanism allows ion acceleration to GeV energies. 
Using the flying focus geometry illustrated in Fig.~\ref{fig:PIC_2d_two_pulse}a, a laser with total energy $0.8\,$kJ and duration $21\,$ps focused in a hydrogen plasma with electron density $n_e \approx 5.5\times10^{19} $cm$^{-3}$ produced the fields shown in Fig.~\ref{fig:PIC_3d_en_theta}a.
The laser power was $37.5\,$TW and the peak intensity at the focal position was $I_0\approx3\times10^{20}\,\mathrm{W/cm^{2}}$, comparable to the capabilities of contemporary large-scale laser facilities.
After accelerating for $6400 T_0$ and propagating a distance of $4.4\,$mm, the protons attained the angular-energy distribution shown in Fig.~\ref{fig:PIC_3d_en_theta}b.  
The proton beam was collimated along the longitudinal direction with divergence angle  $\overline{\theta}_i\approx0.61^\circ$ and exhibited a quasi-monoenergetic spectrum peaked at $\mathcal{E}_i^{peak}\approx 1.64\,$GeV with $\delta\mathcal{E}_i/\mathcal{E}_i^{peak}\approx 3.70\%$ relative energy spread.
The total charge of the proton beam was $Q\approx 23.1\,$pC.
The correlation between the proton energy $\mathcal{E}_i$ and the acceleration distance $x$ for different values of $\Pi_0$ in Fig.~\ref{fig:PIC_3d_en_theta}c demonstrates that the protons were synchronously accelerated by the charge-separation field produced by the flying-focus laser pulse.

To explain and generalize these results, an analytic theory was developed based on a conserved Hamiltonian $\mathcal{H}$ for the ion dynamics in the co-moving frame of the flying focus.
In the co-moving frame, ions evolve according to $d\xi/dt = p/(m_i\gamma)-\Tilde{p}/(m_i\Tilde{\gamma})$ and $d\dot{\xi}/dt = (eZ_i/m_i)[E(\xi)/\gamma^3-\Tilde{E}/\Tilde{\gamma}^3]$, where $\xi\equiv x-x_f$ is the relative coordinate, $\Tilde{p}=\Pi_0 m_i t$, $\Tilde{E}=m_i\Pi_0/(eZ_i)$, and $\Tilde{\gamma}=[(\Pi_0t/c)^2+1]^{1/2}$. 
The ion motion follows a conserved Hamiltonian~\cite{Supplemental_Material}
\begin{eqnarray}\label{eq:hami}
\small
\mathcal{H}(\xi, \dot{\xi} )=\frac{1}{2}m_i\dot{\xi}^2 + \frac{eZ_i}{\Tilde{\gamma}^3}\Phi(\xi)+\frac{m_i\xi\Pi_0}{\Tilde{\gamma}^3},
\end{eqnarray}
where $\Phi(\xi)=-\int E(\xi) d\xi$ is the electric potential. Ions can be efficiently accelerated to relativistic energy if they are trapped inside the $(\xi,\dot{\xi})$ separatrix (Fig.~\ref{fig:Hami_inertial}a), where $\xi\equiv x-x_f$ is the coordinate in the co-moving frame and the overdot denotes a time derivative.
From this, the criterion for ion trapping is $\Pi_0<\Pi_*\equiv eZ_i E_\textrm{max}/m_i$, which sets a relationship between the acceleration $\Pi_0$ and the maximum strength of the accelerating electric field $E_\textrm{max}$. 
The trapping criterion corresponds to the appearance of a local minimum in the total potential energy $\Psi=eZ_i\Phi+\Pi_0 m_i\xi$ (Fig.~\ref{fig:Hami_inertial}a), which is the sum of the electric energy $eZ_i\Phi$ and the inertial energy from the co-moving frame $\Pi_0 m_i\xi$.
The upper and lower panels of Fig.~\ref{fig:Hami_inertial}a show large and small inertial potentials, respectively. A potential valley appears for the total potential $\Psi$ that satisfies the trapping criterion; ions within this valley oscillate longitudinally in $\xi$, which is equivalent to rotating within the separatrix.
The left and right boundaries of the potential valley ($\xi_{s,l}$ and $\xi_{s,r}$) have equivalent points on the separatrix at $\dot{\xi}=0$. 
The local minimum potential at $\xi=\xi_f$, where the electric field and the inertial force are balanced, corresponds to the fixed point $(\xi_f,0)$ in $(\xi,\dot{\xi})$ phase space where the ion is static in the comoving frame.
The separatrix divides $(\xi,\dot{\xi})$ phase space into trapping and sliding regimes. The ions initialized with $\xi_{s,l}<\xi|_{t=0}<\xi_{s,r}$ at $\dot{\xi}=0$ are trapped and efficiently accelerated while those with $\xi|_{t=0}<\xi_{s,l}$ or $\xi|_{t=0}>\xi_{s,r}$ slide away from the electric field.
For increased $\Pi_0$, the potential valley shrinks. At the critical condition where  $\xi_{s,l}= \xi_f = \xi_{s,r}$, the separatrix vanishes and no ions can be accelerated.

Particle trajectories found in 2D PIC simulations match this analysis~\cite{animation_fig4_particle_tracking}, as shown in Fig.~\ref{fig:Hami_inertial}b-e, where Fig.~\ref{fig:Hami_inertial}bc and Fig.~\ref{fig:Hami_inertial}de correspond to  $\Pi_0=1.2\times 10^{-3}cT_0^{-1}>\Pi_*$ and $\Pi_0=4\times10^{-4}cT_0^{-1}<\Pi_*$, respectively. Here the threshold for separatrix disappearance is $\Pi_*\approx 7.8\times10^{-4}cT_0^{-1}$, so only sliding proton dynamics exist in the case of higher acceleration. Protons were trapped and accelerated in the lower acceleration case, as shown by the momentum accumulation (Fig.~\ref{fig:Hami_inertial}d) and the separatrix in phase space (Fig.~\ref{fig:Hami_inertial}e).
Figure~\ref{fig:Hami_inertial}f shows the time evolution of a trapped proton trajectory in the co-moving frame. 
The average acceleration gradient $d\mathcal{E}_i/dx\approx 4.0\,$GeV/cm agrees well with the acceleration provided by the drifting electric field $\Pi_0m_i\approx4.1\,$GeV/cm. Microscopically, the proton undergoes multiple cycles of nonuniform acceleration within the field, corresponding to oscillations in the separatrix shown in Fig.~\ref{fig:Hami_inertial}e.

\begin{figure}
\centering
\includegraphics[width=0.48\textwidth]{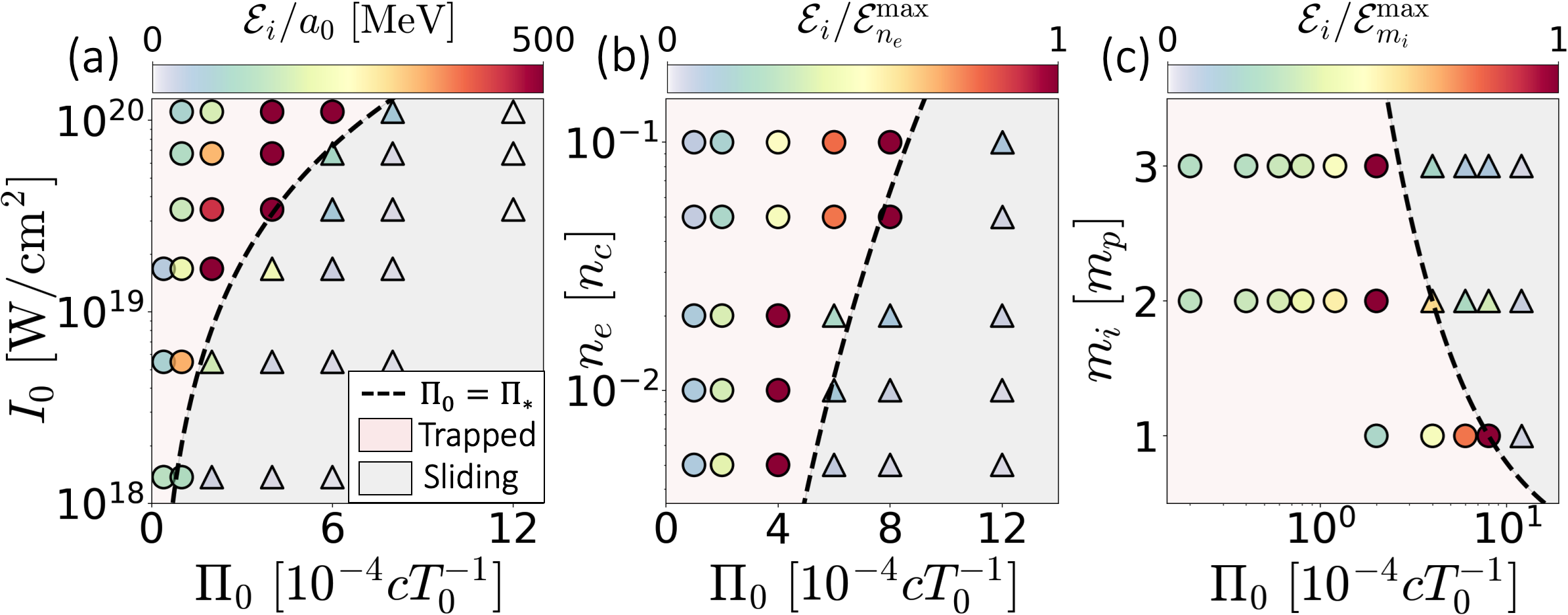}
\caption{2D PIC simulations showing ion energy as a function of laser intensity $I_0$,  flying focus acceleration $\Pi_0$, plasma density $n_e$, and ion mass $m_i$.
(a) Intensity and acceleration are varied. In (b) plasma density and in (c) ion mass is varied. For all panels, unspecified parameters are $a_0=9$, $\sigma_0=4$~$\upmu$m, $m_i=m_p$, and $n_e=0.05n_c$. Black dashed lines mark the predicted threshold for ion trapping $\Pi_0 = \Pi_{*}$. The grey region corresponds to the sliding regime with $\Pi_0>\Pi_{*}$.
Symbol colors indicate maximum achieved energy, with circles and triangles indicating trapping and no trapping, respectively.
}
\label{fig:PIC_2d_para_scan}
\end{figure}

An explicit formula for the trapping criterion $\Pi_*$ was found by calculating the self-generated electrostatic field $\overline{E}_x$ at the laser-driven plasma channel edge. The field was estimated by balancing the average ponderomotive force of the laser pulse in the $x$-direction $F_{p,x}$, calculated as $\overline{F}_p=\int_0^{2\sigma_0}F_{p,x}dx/(2\sigma_0)\sim a_0^2m_ec^2/(\overline{\gamma}_e \sigma_0)$, with the electrostatic force $-e\overline{E}_x$ exerted on plasma electrons. Here $a_0$ is the normalized vector potential, $\sigma_0$ is the laser focal spot size, \re{and $\overline{\gamma}_e$ is the relativistic Lorentz factor of expelled electrons, where the relation of $\overline{\gamma}_e\sim a_0 (n_e/n_c)^{-1/6}$ can be obtained from the numerical fitting of PIC simulation results.}
Utilizing the electric field $e\overline{E}_x\sim\overline{F}_p$ and $\overline{E}_x\sim a_0(n_e/n_c)^{1/6}m_ec^2/(e\sigma_0)$,
the threshold for ion trapping by the accelerating field is:
\begin{eqnarray}\label{eq:Pi_th}
\Pi_0 <\Pi_{*}\sim \frac{Z_im_e c^2a_0(n_e/n_c)^{1/6}}{m_i\sigma_0}.
\end{eqnarray}
This trapping threshold ($\Pi_{*}$) was validated with 2D PIC simulations across different laser intensities, focal spot sizes ($\sigma_0$), plasma electron densities ($n_e$), and accelerations $\Pi_0$.
The analytic threshold shown in Fig.~\ref{fig:PIC_2d_para_scan} (dashed lines) agrees well with the simulations whether the ions are trapped and accelerated (circles) or slip without acceleration (triangles).
In Fig.~\ref{fig:PIC_2d_para_scan}a, the threshold $\Pi_{*}$ increases with rising laser intensity because the plasma self-generated electric field $\overline{E}_x$ is proportional to $a_0$. 
As the accelerating field $\overline{E}_x$ is positively related to the plasma electron density $n_e$, the trapping threshold $\Pi_{*}$ rises with increasing $n_e$ (Fig.~\ref{fig:PIC_2d_para_scan}b).
Heavier ions are more difficult to accelerate, resulting in a smaller trapping threshold $\Pi_{*}$ for larger ion masses (Fig.~\ref{fig:PIC_2d_para_scan}c).
As shown in each of these subfigures, accelerations close to, but below, the threshold produce the highest ion energies. 

In experiments, imperfections and inaccuracies in the optical elements could lead to fluctuations in the spatiotemporal profile of the flying-focus pulse. 
Additional PIC simulations have been performed to demonstrate that moderate fluctuations do not influence the efficiency of the LWFA-TFF mechanism~\cite{Supplemental_Material}. These simulations demonstrate that the mechanism is ensured by the condition of a Hamiltonian separatrix emerging in $(\xi,\dot{\xi})$ space as elucidated in the analyses above.
While previous experiments have verified the flying focus at lower laser intensities~\cite{froula2018spatiotemporal,turnbull2018raman,jolly2020controlling,kabacinski2023spatio,fu2024steering,liberman2024use,pigeon2024ultrabroadband},
the latter two of these~\cite{liberman2024use,pigeon2024ultrabroadband} demonstrated techniques can be applied at the greater than $ 10^{19}\,\mathrm{W/cm^2}$ intensities required for LWFA-TFF.
These developments indicate the possibility of experimentally achieving the TFF optical configuration in the near future.

The application of a transverse flying focus to ion acceleration offers a potential path towards $>$100 MeV quasi-monoenergetic ions from laser-driven sources. 
Accelerating ions in an underdense plasma removes several limitations of TNSA and other ion-acceleration mechanisms that require overdense targets, including deformation due to the laser prepulse~\cite{kaluza2004influence}, difficulty accelerating heavy ions in the presence of protons \cite{shen2017achieving_PRL,wang2021super,martin2024narrow}, and complex and low-repetition rate targetry \cite{higginson2018near_94MeV,ma2019laser,dover2023enhanced}, making this method potentially suitable for high-dose applications.
The use of lower laser intensities and plasma densities may also mitigate kinetic instabilities \cite{palmer2012rayleigh,wan2020effects}, avoid laser energy dissipation at ultrarelativistic intensities \cite{di2012extremely,ji2014radiation,gonoskov2022charged}, and relax requirements on the laser facility.
Like LWFA of electrons, LWFA-TFF produces quasi-monoenergetic, collimated, and high-energy particle beams. 

In 3D simulations, the conversion efficiency of laser energy into the entire proton population is $\eta\approx 0.13\%$. However, the traditional conversion efficiency $\eta$ does not reflect the key properties of an accelerated proton beam for applications that require monochromatic energy and collimation. Here, LWFA-TFF has an advantage in the effective conversion ratio $\eta^* \equiv E_{p,1\%}/(E_l\Omega)\sim 10^{-4} \mathrm{msr}^{-1}$ of laser energy $E_l$ to a collimated monochromatic ion beam~\cite{Supplemental_Material}, where $E_{p,1\%}$ is the proton energy within 1\% spread of the peak energy and $\Omega$ is the solid angle of the beam divergence (msr means Millisteradian).

The transverse flying focus and its implementation in a ring focusing geometry offer new ways to control and apply high-strength electromagnetic fields that may have applications beyond ion acceleration. In particular, the complete decoupling of focal spot motion from the direction of light propagation may also allow improvements to electron acceleration, photon acceleration, or terahertz radiation generation. 
Here, it enables the efficient and stable acceleration generation of monoenergetic ion beams in the critical 0.1--1 GeV/nucleon energy range with laser parameters that are accessible today, while avoiding complex targets, exquisitely tuned plasma parameters, and extreme intensities, creating a potential new route to the widespread application of compact laser-based ion sources.

\begin{acknowledgments}
This work was partially supported by NSF Grant PHY-2308641 and NNSA Grant DE-NA0004130. 
The work of JPP is supported by the Office of Fusion Energy Sciences under Award Number DE-SC0021057, the University of Rochester, and the New York State Energy Research and Development Authority. The PIC code EPOCH~\cite{arber2015contemporary} is funded by the UK EPSRC grants EP/G054950/1, EP/G056803/1, EP/G055165/1 and EP/ M022463/1. This work used Delta-cpu at National Center for Supercomputing Applications (NCSA) through allocation PHY230120 and PHY230121 from the Advanced Cyberinfrastructure Coordination Ecosystem: Services \& Support (ACCESS) program, which is supported by National Science Foundation grants \#2138259, \#2138286, \#2138307, \#2137603, and \#2138296. 
We would like to thank Stanford University and the Stanford Research Computing Center for providing computational resources of (Stanford) Sherlock cluster.

\end{acknowledgments}

\input{output.bbl}

\end{document}

%% file: output.bbl
%